\documentclass{article}

\usepackage{arxiv}

\usepackage[utf8]{inputenc} 
\usepackage[T1]{fontenc}    
\usepackage{hyperref}       
\usepackage{url}            
\usepackage{booktabs}       
\usepackage{amsfonts}       
\usepackage{nicefrac}       
\usepackage{microtype}      
\usepackage{lipsum}
\usepackage{graphicx}
\usepackage{multirow}
\usepackage{amsmath}

\title{Evaluating the Effect of Pretesting with Conversational AI on Retention of Needed Information}

\author{
  Mahir Akgun \\
  College of Information Sciences and Technology\\
  Pennsylvania State University\\
  University Park, PA 16827 \\
  \texttt{makgun@psu.edu} \\
   \And
 Sacip Toker \\
  Information Systems Engineering\\
  Atılım University\\
  Ankara, Türkiye \\
  \texttt{sacip.toker@atilim.edu.tr} 
  \\
}

\begin{document}
\maketitle
\begin{abstract}
This study explores the role of pretesting when integrated with conversational AI tools, specifically ChatGPT, in enhancing learning outcomes. Drawing on existing research which demonstrates the benefits of pretesting in memory activation and retention, this experiment extends these insights into the context of digital learning environments. A randomized true experimental study was utilized. Participants were divided into two groups: one engaged in pretesting before using ChatGPT for a problem-solving task involving chi-square analysis, while the control group accessed ChatGPT immediately. The results indicate that the pretest group significantly outperformed the no-pretest group in a subsequent test, which suggests that pretesting enhances the retention of complex material. This study contributes to the field by demonstrating that pretesting can augment the learning process in technology-assisted environments by preparing the memory and promoting active engagement with the material. The findings also suggest that learning strategies like pretesting retain their relevance in the context of rapidly evolving AI technologies. Further research and practical implications are presented. 
\end{abstract}


\section{Introduction}
In the digital age, where information is readily accessible with a single search, the role of search engines in facilitating learning experiences has become increasingly prominent. Traditional approaches to studying often involve passive consumption of information, but recent research has suggested that actively engaging with the material through methods such as pretesting—where participants are asked to retrieve information from memory before studying it—could significantly enhance learning outcomes. Pretesting engages participants in active retrieval processes, not only assessing current knowledge but also priming the memory for subsequent encoding of related information. In essence, pretesting serves as an initial assessment or diagnostic tool before formal learning occurs and can strengthen memory traces and facilitate long-term retention.

Leveraging the ubiquitous nature of search engines, particularly Google, researchers have explored the potential benefits of pretesting in improving subsequent recall for searched-for content. While search engines provide instant access to vast amounts of information, questions have arisen regarding the depth of understanding and retention achieved through such methods. In a study by Giebl et al. \cite{giebl2021answer}, some but not all of the participants (the pretest vs. no-pretest groups) were asked to solve a challenging problem before consulting Google for needed information. The results revealed that this pretesting method significantly enhanced participants' subsequent recall of the searched content. These findings highlighted the potential of pretesting as a search-for-learning strategy to deepen learning and improve memory retention. 

Building upon these findings, the present study aims to extend the investigation into the efficacy of pretesting as a learning strategy by using conversational AI (CAI) as the information retrieval tool. Specifically, we seek to examine whether pretesting, when implemented with a CAI model, such as ChatGPT, can yield similar enhancements in participants' subsequent recall of the searched content. By exploring the potential of ChatGPT as a learning aid, we aim to contribute to a deeper understanding of optimal learning strategies in the age of AI.

\section{Background}
\label{sec:headings}
\subsection{The Benefits of Pretesting}
Pretesting is a pedagogical strategy where learners are exposed to questions or challenges prior to engaging with learning materials. This method is foundational for enhancing learning and memory recall by activating relevant cognitive frameworks and motivating active engagement with new information \cite{wissman2011interim}. By setting the stage for learning, pretesting helps integrate new knowledge more effectively and sustainably.

In exploring the benefits of pretesting, Carpenter and Toftness \cite{carpenter2017effect} specifically utilized prequestions, which are a form of pretesting that involves presenting learners with specific questions before they encounter educational content. Their findings revealed that prequestions before video segments significantly improved retention of both questioned and non-questioned material, suggesting that this form of pretesting can guide attention and foster deeper processing of information. Similarly, Carpenter et al. \cite{carpenter2018effects} implemented prequestions in a classroom setting and demonstrated that this approach effectively focuses student learning on key concepts.

The value of pretesting extends beyond guided questioning. Richland et al. \cite{kornell2009unsuccessful} highlighted that even unsuccessful retrieval attempts during pretesting can enhance learning, potentially by directing attention toward crucial information and fostering deeper engagement during subsequent study. This was further supported by Kornell et al. \cite{kornell2009unsuccessful}, who reassured educators that unsuccessful tests, especially with subsequent feedback, are beneficial rather than detrimental, thus endorsing the use of pretests as learning events. Grimaldi and Karpicke \cite{grimaldi2012and} explored how semantic relationships influence the efficacy of pretesting, finding that retrieval attempts prior to study are more beneficial when materials are semantically related. This supports the search set theory, where related knowledge activation aids new information encoding. Additionally, Kornell \cite{kornell2014attempting} addressed the timing of feedback in pretesting, which reveals that being asked a question and failing to provide a correct answer can positively influence learning, even when the correct answer is not revealed until a significant delay afterward. These results suggest that characteristics of the learning materials play a critical role in shaping the impact of delayed feedback subsequent to an incorrect response.

Overall, the research on pretesting supports its significant role in enhancing academic performance and deepening learners' understanding across diverse educational settings. By preparing learners mentally for new information, focusing their attention, and facilitating deeper engagement with the material, pretesting proves to be an indispensable strategy for improving learning outcomes and memory retention.

\subsection{Pretesting and Internet}
The scope of information accessible via the internet far exceeds what any human partner can store; it offers a more straightforward and quicker method for accessing current information. At times, retrieving information from the internet can be faster than recalling it from our own memory. Consequently, the distinction between the internal mind (i.e., what is stored in our memory) and the external mind (i.e., what is known by a partner) diminishes when we form a transactive memory system with the internet.

Individuals can delegate more information to search tools, which act as their memory partner, than they would retain internally. This reliance on such a memory partnership shifts our focus from what information to remember to where information can be found, thereby easing cognitive load \cite{risko2016cognitive}. In this partnership, the Internet holds all the knowledge while the human partner leverages it for effortless external retrieval \cite{fisher2015searching}. This process may result in bypassing potentially productive processes triggered by pretesting that potentiate learning of new information.

However, the research by Giebl et al. \cite{giebl2021answer} suggests that integrating pretesting strategies into our use of the internet for information retrieval can enhance the effectiveness of such human-internet partnerships. Their study found that attempting to answer questions or solve problems before consulting the internet not only improved subsequent recall of the searched-for content but also enhanced the retention of previously studied information. Particularly for individuals with some prior knowledge of the topic, pretesting before using Google to find answers resulted in better performance on later tests, suggesting that activating prior knowledge helps integrate new information in ways that are beneficial for learning. This finding underscores the potential of pretesting as a cognitive strategy that not only utilizes external digital resources more effectively but also promotes deeper engagement and retention. In line with these observations, Giebl et al. \cite{giebl2023thinking} further elaborates on the cognitive benefits of "thinking before googling" by demonstrating that engaging with a question from memory before seeking answers online leads to significantly better recall and learning outcomes compared to immediately googling for answers. This active engagement prepares the memory, effectively leveraging the internet as a tool that complements rather than supplants personal cognitive efforts.

\section{Present Study}
The integration of advanced conversational AI tools like ChatGPT into learning environments represents a significant shift in how we interact with information technology for educational purposes. These tools not only facilitate access to vast amounts of information but also provide a dynamic platform for engaging with content through dialogue and interactive feedback. This presents an opportunity to explore innovative teaching strategies that could complement learning methods.

Building on the demonstrated benefits of pretesting in enhancing information retrieval and retention with internet use, this study aims to investigate if a similar effect can be observed when learners engage with conversational AI. The study examines whether engaging users in a pretesting phase before they interact with ChatGPT can enhance their learning outcomes. The focus is on determining how pretesting might influence the retention of information provided by ChatGPT, especially when users are prompted to use their memory before seeking AI assistance.

This study is crucial for several reasons. First, it seeks to extend the existing knowledge on pretesting in the context of rapidly evolving AI technologies. Second, by exploring how pretesting can be effectively integrated with AI-driven learning tools, this research contributes to our understanding of how educational strategies can be adapted to leverage the strengths of AI tools.

\section{Method}
\subsection{Experimental Design}
The study, designed as a true-experimental posttest with randomized groups \cite{field2002design}, was carried out in an intermediate-level statistics course and consisted of two phases. In the first phase, all participants engaged with materials covering one-way chi-square analysis. This was followed by a second phase, where participants faced a complex cybersecurity problem that required them to conduct a two-way chi-square analysis. The problem introduced in the second phase was deliberately designed to be unsolvable using only the information from the first phase. This approach ensured that while students had a basic understanding of chi-square analysis, their knowledge was insufficient for applying two-way chi-square analysis to more complex problems. This setup allowed students to engage with more complex issues based on their preliminary knowledge, facilitating the establishment of a pretesting task within the study design.

Specifically, participants in the no-pretest group were immediately given access to the necessary information through ChatGPT. Conversely, those in the pretest group attempted to solve the problem independently before they could access this additional help. The study concluded with a delayed multiple-choice test to evaluate not only the retention and comprehension of the learned content but also the participants' ability to apply this knowledge in new situations.

\subsection{Participants}
Seventy-three undergraduate students majoring in information sciences at a large public research university in Pennsylvania, United States, were initially recruited for the study. The cohort comprised 17 women and 56 men. Technical issues, such as internet connectivity and user account problems, along with non-adherence to the instructions, led to the exclusion of 12 participants. Consequently, 61 students completed the study, with 31 (50.8\%) assigned to the no-pretest group and 30 (49.2\%) to the pretest group. The study received approval from the Institutional Review Board (IRB No 00024664).

\subsection{Procedure}
The experiment comprised two phases, as illustrated in Figure 1. Both phases occurred within the same week. The course, which convened twice weekly, allotted 75 minutes per session. Phase 1 occurred during the initial session, focusing on one-way chi-square analysis. Phase 2 unfolded in the subsequent session, starting with a recall test to measure participants' understanding of the concepts covered in Phase 1. Subsequently, participants completed a more complex task requiring the use of two-way chi-square analysis, for which they had not yet received all the necessary information.
During Phase 2, participants were randomly assigned to either a pretest group or a no-pretest group. Those in the pretest group attempted to complete the task before accessing additional information through ChatGPT, while those in the no-pretest group could access the information immediately. This design aimed to evaluate whether attempting the task without full information (pretest) impacted learning compared to accessing all information from the outset (no pretest).
To assess the effects of pretesting on learning, a final multiple-choice test containing questions related to a different but conceptually similar problem scenario was administered to both groups. This test was designed to evaluate participants' understanding of concepts presented in Phase 2.

\begin{figure}[h!]
    \centering
    \includegraphics[width=\textwidth]{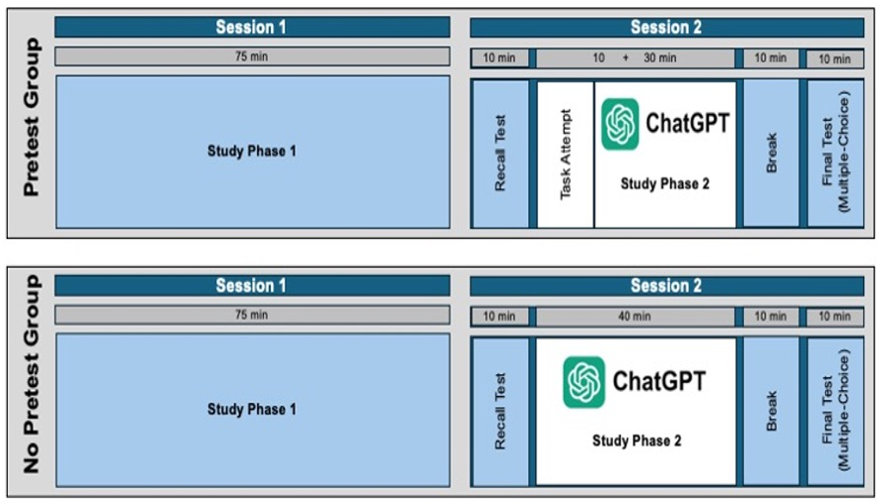} 
    \caption{Experimental procedure} 
    \label{fig:figure1} 
\end{figure}

\subsection{Materials}
\paragraph{Phase 1 Activity.} During Phase 1, participants received instructions on the fundamental aspects of chi-square tests, including (a) the definition of a chi-square test, (b) the rationale behind using chi-square tests, (c) the calculation and interpretation of chi-square values, and (d) the application of one-way chi-square tests to cybersecurity problems. Following the instructions, participants collaborated in groups to apply their knowledge by solving a practical problem (see Figure 2).

\begin{figure}[h!]
    \centering
    \includegraphics[width=\textwidth]{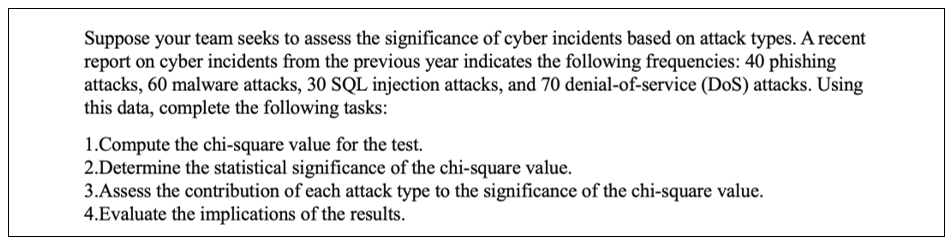} 
    \caption{Scenario used in Phase 1} 
    \label{fig:figure2} 
\end{figure}

\paragraph{Recall Test.}The recall test aimed to assess students' understanding of key concepts related to chi-square tests, as taught in Phase 1. This test was administered to both the No-Pretest Group and the Pretest Group. Participants were presented with a scenario and asked to answer multiple-choice questions based on the scenario (see Figure 3). Two of them are illustrated in Figure 3. The outcomes of this recall test were employed to assess for any statistically significant differences between the No-Pretest Group and the Pretest Group. 

\begin{figure}[h!]
    \centering
    \includegraphics[width=\textwidth]{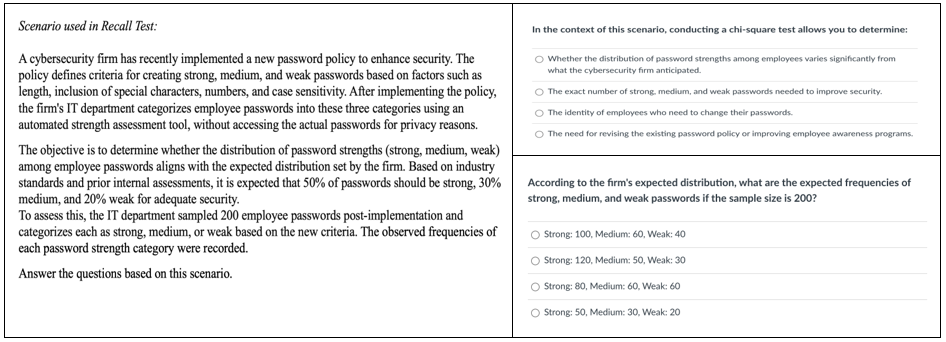} 
    \caption{Scenario and two of the items used in the recall test.} 
    \label{fig:figure3} 
\end{figure}

\paragraph{Phase 2 Activity.} During Phase 2, participants were presented with a cybersecurity problem scenario that required the use of two-way chi-square analysis (see Figure 4). This task built upon concepts studied in Phase 1, such as calculating chi-square values and finding critical values for assessing statistical significance. However, it also required information only available via ChatGPT, such as calculating expected frequencies and determining degrees of freedom for a two-way chi-square test.

Participants assigned to the pretest group did not have immediate access to ChatGPT. Instead, they were instructed to attempt to solve the task without assistance for a period before being given access to ChatGPT. No corrective information was provided during their problem-solving attempts. To ensure that participants engaged cognitively with the problem, participants were asked to answer questions in the pretest (see Figure 4 for pretest questions). Participants who did not provide answers to these questions were excluded from the study. The information necessary to solve the task could be accessed via ChatGPT, which became available to participants in the pretest group once they had submitted their answers to the questions in the pretest.

\begin{figure}[h!]
    \centering
    \includegraphics[width=\textwidth]{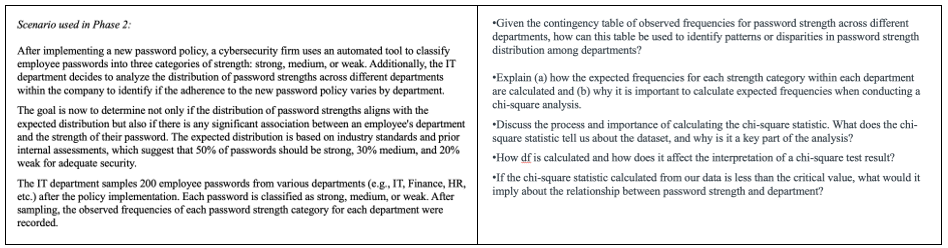} 
    \caption{Scenario used in Phase 2 and pretest questions} 
    \label{fig:figure4} 
\end{figure}

In contrast, participants in the no-pretest group could access ChatGPT immediately to find the relevant information necessary to solve the task. To ensure consistency in the information accessed and to control for potential variations in prompting strategies, participants were provided with sample prompts and comprehensive questions (see Figure 5). These prompts were designed to guide participants through the analytical process step-by-step, ensuring that they engaged with the information in a systematic manner.

Participants were instructed to follow a specific sequence when interacting with ChatGPT:
\begin{enumerate}
    \item \textbf{Introduce the Data:} Begin by clearly communicating the dataset and the problem to ChatGPT using the preliminary step prompt

    \item \textbf{Step-by-Step Engagement:} Progress through each analytical step using the provided direct prompts. This approach will help you methodically understand and apply the chi-square analysis to the dataset.

    \item \textbf{Comprehension and Reflection:} Answer the comprehension questions following each step. These should be in your own words and reflect a deep understanding of the statistical process and its implications.

    \item \textbf{Concluding Reflections:} After completing the analysis, reflect on the significance of your findings and consider their implications in the context of cybersecurity and beyond.
\end{enumerate}

\begin{figure}[h!]
    \centering
    \includegraphics[width=\textwidth]{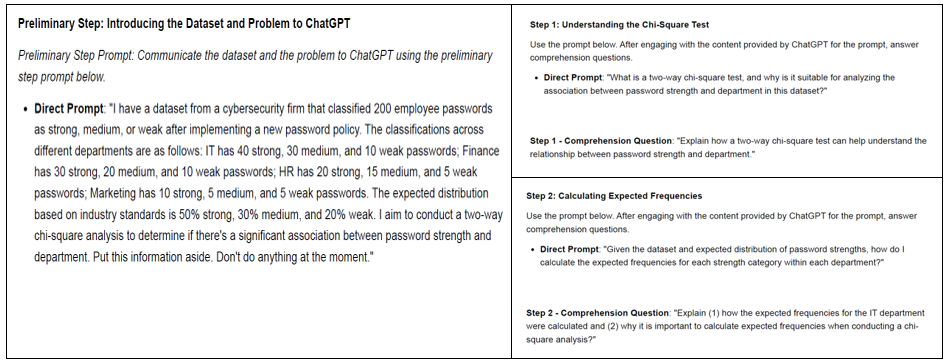} 
    \caption{Preliminary step and first two steps with prompts and comprehensive questions used in Phase 2 activity.} 
    \label{fig:figure5} 
\end{figure}

To prevent direct copying and pasting from ChatGPT in their responses, participants were required to include a copy of their conversation with ChatGPT in their submissions.

\paragraph{Final (Transfer) Test.} The final test was administered to both the no-pretest group and the pretest group. Participants were presented with a scenario and asked to answer multiple-choice questions based on the scenario (see Figure 6). The final test materials consisted of 11 multiple-choice transfer questions on concepts from both study phases, and two of them are illustrated in Figure 6.

\begin{figure}[h!]
    \centering
    \includegraphics[width=\textwidth]{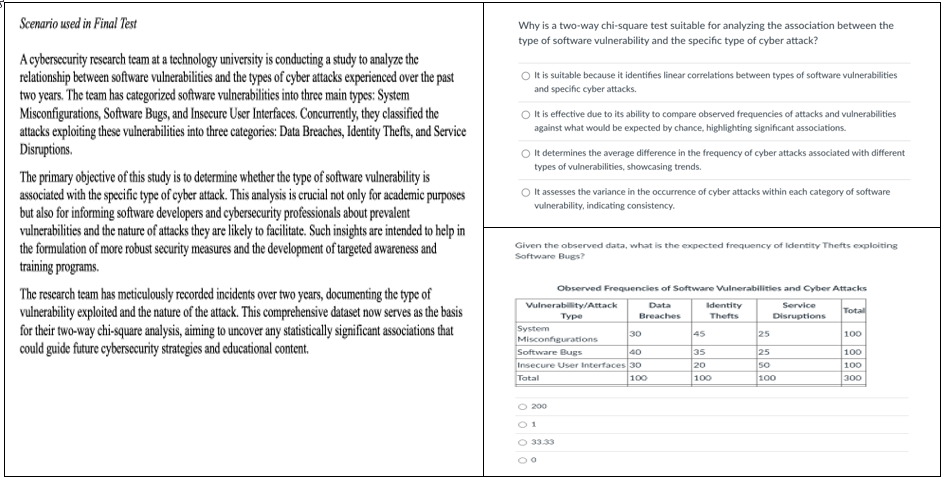} 
    \caption{Scenario and two of the items used in the final test.} 
    \label{fig:figure6} 
\end{figure}

\subsection{Data Analysis}
The normal distribution of recall test and final test scores were examined. The final test results showed skewness (-.059) and kurtosis (-.548) values of $\pm$1.5, indicating no issue \cite{tabachnick2013using}. The recall test results were -1.593 and 2.296, respectively, with one outlier detected in the experimental group. After it was removed, the skewness and kurtosis values were reduced to -1.295 and.657, respectively, which fell below the threshold. As a result, data from 31 participants in the no-pretest group and 29 participants in the pretest group were used in the analysis. 

To determine the potential impact of students' knowledge of chi-square tests prior to Phase 2 activities, we initially considered running an analysis of covariance (ANCOVA) if participants in the pretest group significantly outperformed or underperformed compared to those in the no-pretest group in the recall test \cite{field2017discovering}. Alternatively, we planned to conduct an independent t-test if there was no statistically significant difference in recall test performance between the two groups \cite{field2017discovering}.

\paragraph{Recall Test Performance.} We assessed the recall test performance using an independent t-test, treating the group (pretest group or no-pretest group) as the categorical independent variable and the recall test scores as the continuous dependent variable. The results indicated no significant difference in average scores between the two groups, t(58) = 0.584, p = 0.562). As indicated in Figure 7a, the descriptive statistics for the pretest group were M = 71.03, SD = 13.72, and for the no-pretest group, M = 72.90, SD = 11.01. These findings suggest that participants had a similar level of understanding of the one-way chi-square analysis concepts covered in Phase 1, indicating that this prior knowledge was unlikely to significantly impact the results of the subsequent test related to two-way chi-square analysis.

\begin{figure}[h!]
    \centering
    \includegraphics[width=\textwidth]{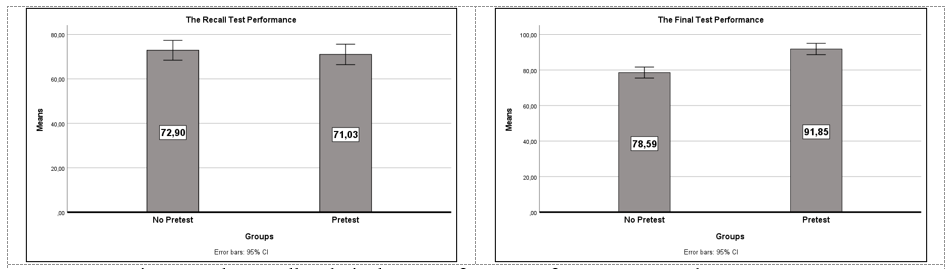} 
    \caption{The Recall and Final Test Performances for No-pretest and Pretest groups} 
    \label{fig:figure7} 
\end{figure}

\paragraph{Final Test Score Analysis.} Based on the similarity in recall test performance between the two groups, we proceeded with an independent t-test. In this test, the students' final test scores were treated as the continuous dependent variable, while the group they were assigned to (pretest group or no-pretest group) was considered the categorical independent variable.

\section{Results}
The independent sample t-test results revealed a significant difference between the pretest and no-pretest groups in the final test scores, t(58) = -5.974, p < .001, Cohen’s d = 8.59, indicating a large effect size \cite{vacha2004estimate}. As shown in Figure 7b, the mean score for the no-pretest group was 78.59 (SD = 7.62), while that of the pretest group was 91.85 (SD = 9.52).

\section{Discussion}
The purpose of our study was to investigate the efficacy of pretesting as a learning strategy when using conversational AI tools like ChatGPT. In our study, we introduced students to one-way chi-square analysis concepts in Phase 1. During Phase 2, participants were asked to tackle a cybersecurity problem requiring two-way chi-square analysis, which could not be solved solely with their initial knowledge. Participants in the pretest group attempted to solve the problem before accessing ChatGPT, while those in the no-pretest group were given immediate access. A final multiple-choice test assessed their understanding and ability to transfer these concepts to new contexts. Our results revealed that participants in the pretest group significantly outperformed those in the no-pretest group, confirming that pretesting fosters deeper learning and retention of statistical concepts.

Our study builds upon a substantial body of research demonstrating the benefits of pretesting before accessing to-be-learned information (e.g., \cite{grimaldi2012and, kornell2009unsuccessful, kornell2014attempting, richland2009pretesting}. It extends this line of research into the context of conversational AI tools like ChatGPT, highlighting their potential to complement pretesting strategies in fostering deeper learning and retention. While several previous studies have reported limited benefits of pretesting, especially in classroom settings (e.g., \cite{carpenter2018effects}), our results show that when integrated with conversational AI, pretesting can produce significant improvements in learners' ability to retain and apply complex concepts (see also \cite{giebl2021answer,giebl2023thinking}).

The current findings come from the study designed to explore the potential benefits of pretesting when learning new information using web-based conversational AI tools. Research has indicated that people tend to rely on the Internet for information storage and access, reducing the degree to which they retain that information internally \cite{marsh2019digital, sparrow2011google}. This reliance on search tools has created a "memory partnership" with the Internet, maximizing the volume of information available and optimizing its retrieval. However, as this partnership deepens, individuals tend to remember where facts can be retrieved online rather than retaining the facts themselves \cite{sparrow2011google}. As reliance on the Internet increases, people often bypass their internal memory for information retrieval. However, in line with Giebl et al. \cite{giebl2021answer}, our findings suggest that pretesting might have the potential not only to leverage external digital resources more effectively but also to foster deeper engagement and retention. 
One explanation for the pretesting effect may relate to its role in helping learners evaluate their knowledge accurately (see \cite{giebl2021answer}). Access to web search tools often inflates individuals' confidence in their cognitive abilities, resulting in an overestimation of what they know \cite{ward2013one, hamilton2018blurring}. Cognitive self-esteem (CSE), one's perception of their ability to think and recall information, is often reported to be higher for those with Internet access than those without. For instance, individuals who used Google to answer trivia questions reported significantly higher CSE scores than those without access \cite{ward2013one}. Similarly, Hamilton and Yao \cite{hamilton2018blurring} found higher CSE scores for those who used Google than for those in the no-Google or control conditions. Incorporating pretesting with conversational AI tools like ChatGPT may help learners assess their knowledge more accurately, potentially encouraging them to engage with their internal cognitive resources before seeking external assistance.

Another potential explanation for our findings lies in the design of our study. Participants in the pretest group initially studied information related to chi-square analysis and then attempted to solve a challenging problem requiring the use of a two-way chi-square test without access to ChatGPT.  The lack of prior knowledge could have made the process time-consuming, but Phase 1 of the study formally introduced domain concepts that, while necessary, were insufficient to solve the problem presented in Phase 2. This deliberate sequencing may have helped students deeply engage with the problem-solving task without overburdening their working memory. Such productive engagement aligns with the concept of 'productive failure' \cite{kapur2010productive}, achieved by a study design where an initial exploration phase prompts students to use their prior knowledge to develop approximate solutions to novel problems, followed by a more formal instruction phase with lectures or practice \cite{kapur2012designing}. The exploration phase in productive failure is deliberately structured to challenge students in a way that often leads to initial failure. This approach provides opportunities for students to activate, differentiate, and refine their existing and intuitive knowledge, which is crucial for solving complex problems \cite{sinha2021problem}. Such a design aligns with our findings, which suggest that pretesting enhances learning by preparing the memory and fostering deeper engagement with the material.

Overall, these findings suggest that incorporating pretesting strategies, even in the context of using external resources like ChatGPT, can enhance learning outcomes and improve students' ability to apply their knowledge in new contexts.
Our study demonstrates some implications for educational strategies, particularly through the integration of pretesting with teaching methods. We recommend that educators encourage students to actively engage with their internal resources before turning to external aids such as conversational AI. This practice not only encourages reliance on their own knowledge and skills but also fosters greater engagement and enhances problem-solving capabilities. Moreover, we encourage students to strive to cope with problem-solving tasks independently before seeking external assistance. This practice is likely to solidify their foundational knowledge and enhance retention. By promoting self-reliance in learning, students may develop stronger cognitive skills and achieve a more profound understanding of the material, which could lead to more effective and lasting educational outcomes.
While our results offer promising insights into the benefits of pretesting, they also highlight certain constraints. Our research was conducted within an intermediate-level statistics course in information sciences, which may influence the applicability of the results to other disciplines or levels of education. To expand the generalizability of our findings, it is essential to explore how pretesting influences learning outcomes in different educational fields.

Building on the insights gained, future research should focus on several key areas. Firstly, tailoring pretesting strategies to individual learner profiles could significantly enhance their effectiveness. Considering variations in students' prior knowledge when designing pretesting approaches may help maximize learning outcomes and ensure that these strategies are more effectively adapted to individual needs. Additionally, investigating the long-term effects of pretesting on learners' retention and their ability to practically apply the knowledge acquired is crucial. Understanding how pretesting can influence long-term memory and practical application will provide deeper insights into the pedagogical benefits of this approach and guide the development of more effective educational practices.

\bibliographystyle{unsrt}  


\bibliography{references}
\end{document}